\documentclass[twocolumn,aps,showpacs,prb,tightenlines,amsmath,amssymb]{revtex4}
\usepackage{bm}
\usepackage{graphicx}             
\usepackage{amssymb}               
\usepackage{amsmath}                     
\usepackage{colordvi}
\usepackage{calrsfs} 
\newcommand{\bgreek}[1]{\mbox{\boldmath$#1$\unboldmath}}
\begin{document}   

\title{Electron spin relaxation in rippled graphene with low mobilities}
 
\author{P. Zhang}
\author{Y. Zhou}
\author{M. W. Wu}
\thanks{Author to whom correspondence should be addressed}
\email{mwwu@ustc.edu.cn.}
\affiliation{Hefei National Laboratory for Physical Sciences at
  Microscale and Department of Physics, 
University of Science and Technology of China, Hefei,
  Anhui, 230026, China} 
\date{\today}

\begin{abstract} 

We investigate spin relaxation in rippled graphene where  curvature
induces a Zeeman-like spin-orbit coupling with opposite 
effective magnetic fields
along the graphene plane in ${\bf K}$ and ${\bf K}^\prime$
valleys. The joint effect of this Zeeman-like spin-orbit coupling 
and the intervalley electron-optical phonon scattering opens a spin
relaxation channel, which manifests itself in  low-mobility samples
with the electron mean free path being smaller than the ripple size.
 Due to this spin relaxation
channel, with the increase of temperature, the relaxation
 time for spins perpendicular to the
effective magnetic field first decreases and then increases, 
with a minimum of several
hundred picoseconds around room temperature. However, the spin
relaxation along the effective magnetic field is determined by
the curvature-induced Rashba-type spin-orbit coupling,
leading to a temperature-insensitive spin
relaxation time of the order of microseconds.
 Therefore, the in-plane spin relaxation in
low-mobility rippled graphene is anisotropic. Nevertheless, in
the presence of a small perpendicular magnetic field, as usually
applied in the Hanle spin precession measurement, the anisotropy of spin
relaxation is strongly suppressed.

\end{abstract}
%81.05.ue  Graphene
%71.70.Ej  Spin-orbit coupling, Zeeman and Stark splitting, Jahn-Teller effect
%75.70.Tj  Spin-orbit effects
%85.75.-d  Spintronics

\maketitle
\section{Introduction}
In recent years, graphene has attracted much interest due to its 
potential for the all-carbon based electronics and
spintronics.\cite{novoselov666,Tombros_07,Geim_07,castro,peres2673,sarma407,ep-review,wu-handbook,FWang,beenakker,abergel,shiraishia,acik,geim2011} 
A number of experiments on spin relaxation in graphene on SiO$_2$
substrate have been carried out, with spin relaxation times $\tau_s$ of the order of
10-100~ps reported.\cite{Tombros_07,han222109,Pi,Jozsa_09,Tombros_08,Popinciuc,han1012.3435,jo,avsar}    
Some works suggested the Elliott-Yafet (EY)\cite{ey} mechanism to be
dominant in spin relaxation,\cite{Jozsa_09,Popinciuc,Tombros_08,han1012.3435,avsar,jo}
while the surface chemical doping experiment supported the importance of the
D'yakonov-Perel' (DP)\cite{dp} one.\cite{Pi} These experiments have triggered
intensive theoretical studies on spin relaxation in 
graphene.\cite{hernando146801,Castro_imp,yzhou,Fabian_SR,dugaev085306,arxiv1107.3386,zhanggraphene,jeong,zhangdpey}
With inversion asymmetry, possibly caused by a perpendicular electric
field or curvature, a Rashba-type spin-orbit coupling
(SOC),\cite{Rashba} which couples spin to pseudospin, arises in
graphene.\cite{Kane_SOC,Min,Fabian_SOC,Hernando_06,jeong,abde} It was
found that with this SOC, the DP mechanism dominates
the electron spin relaxation.\cite{hernando146801}  
Besides, it was revealed that for the EY mechanism,
$\tau_s/\tau_p\sim n_e$, where $\tau_p$ is the momentum relaxation time and
$n_e$ is the electron density.\cite{arxiv1107.3386,hernando146801}
This makes the assessment, 
which attributes the observed linear relation between spin relaxation
time $\tau_s$ and momentum relaxation time $\tau_p$ with the increase
of electron density $n_e$ to the EY mechanism,
in the experimental works\cite{Jozsa_09,Popinciuc,Tombros_08,han1012.3435,avsar}
 inappropriate.\cite{jo} The magnitude of the Rashba-type SOC
caused by a moderate perpendicular electric
field\cite{Fabian_SOC,abde} or curvature\cite{Hernando_06,jeong} is of 
the order of $\mu$eV and the corresponding $\tau_s$ limited by the DP
mechanism is as large as microseconds.\cite{yzhou} The 
adatoms were suggested to enhance the 
local SOC to the order of 10~meV and hence provide a possible origin
of the observed short spin relaxation time.\cite{Castro_imp,abde,varykhalov,zhanggraphene,zhangdpey,Fabian_SR}
Recently, we set up a random Rashba model incorporating the effect of
adatoms\cite{zhangdpey} and fitted the experimental data from
different groups.\cite{Jozsa_09,Pi,han1012.3435,jo} We suggested that the DP mechanism dominates spin
relaxation in graphene and can result in either linear or inversely
linear relation between $\tau_s$ and $\tau_p$.\cite{zhangdpey}

Very recently, Jeong {\it et al.} reported that curvature in graphene 
can lead to not only the Rashba-type SOC which is off-diagonal in the
pseudospin space, but also an
additional SOC diagonal in the pseudospin space.\cite{jeong}
This additional SOC serves as a 
Zeeman-like term with opposite magnetic fields in ${\bf K}$ and ${\bf
  K}^\prime$ valleys, similar to the case in carbon
nanotube.\cite{izumida,jeong075409,semenov} Starting from the effective SOC, Jeong {\it et al.} studied
spin relaxation in a chemically-clean corrugated graphene where the
electron mean free path is much larger than the spatial range of
the random spin-orbit fluctuation 
 and the spatially averaged SOC is zero. Under such condition, the spin
relaxation is solely limited by the spin-flip
scattering due to the fluctuation of the SOC,\cite{glazov2157,sherman67,dugaev085306} as in the rippled graphene studied by
Dugaev {\it et al.} where only the Rashba-type SOC was
considered.\cite{dugaev085306} It is noted that in the two valleys, even though the
effective magnetic fields from the Zeeman-like term are opposite, their contributions to the spin relaxation
  via the fluctuation-induced spin-flip scattering are independent and
  identical. Besides, it was found that the
  fluctuations of the Rashba-type SOC and the Zeeman-like term  make comparable contribution
  to the spin relaxation, leading to a spin relaxation time at least of
  the order of 10~ns.\cite{jeong}

Nevertheless, for low-mobility samples where the electron mean
  free path  is smaller than the spatial range of the random
  spin-orbit fluctuation, the spin relaxation is determined by the {\em local}
curvature-induced SOC.\cite{yzhourandom} For each isolated valley, the Rashba-type SOC, together with the Zeeman-like term, still
leads to a spin relaxation time of the order of microseconds in the DP mechanism. However, the
two valleys are not independent any more due to the intervalley scattering. The opposite effective magnetic fields in two valleys,
together with the intervalley scattering, give rise to another
spin relaxation channel. This can be understood by a simple model. We label the spin vector in each valley as ${\bf S}_\mu$, where
$\mu=\pm 1$ stands for the valley located at ${\bf K}$ or ${\bf
  K}^\prime$ point. The spin vector in each valley precesses around
the effective magnetic field from the Zeeman-like term with a
frequency ${\bgreek \omega}_\mu$, where ${\bgreek
  \omega}_\mu=-{\bgreek \omega}_{-\mu}$ and $|{\bgreek
  \omega}_\mu|=\omega$. The intervalley scattering is
characterized by a scattering time $\tau_v$ between two valleys. The
spin vectors ${\bf S}_\mu$, with the initial values ${\bf
  S}_\mu(0)={\bf S}_{-\mu}(0)={\bf S}_0$, satisfy the rate equations
\begin{equation}
\dot{\bf S}_\mu(t)+{\bf S}_\mu(t)\times {\bgreek \omega}_\mu+[{\bf
S}_\mu(t)-{\bf S}_{-\mu}(t)]/\tau_v=0.
\label{simp-analy}
\end{equation}
When ${\bf S}_0$ lies in the plane perpendicular to the effective
magnetic field, one has the solution 
\begin{equation}
\sum_\mu{\bf S}_\mu(t)=2{\bf S}_0e^{-t/\tau_s}
\label{s1}
\end{equation}
with $\tau_s=2/(\omega^2\tau_v)$  in the strong intervalley scattering limit $\omega\tau_v\ll 1$.
So far this mechanism has not been revealed in the literature.

In this work, we study the spin relaxation in the low-mobility rippled 
graphene (the mobility is around $2\times
10^3$~cm$^2$/V$\cdot$s)\cite{avsar} and take into account the above spin
relaxation channel. The electron mean free path $l$ is smaller than
  the ripple size $\xi$.
 In the low temperature regime where the intervalley electron-phonon
scattering is negligible, the spin relaxation is dominated
by the Rashba-type SOC and $\tau_s$ is as large as microseconds. However, with the increase of temperature, due to the above spin relaxation
channel, the relaxation time for spins polarized perpendicular to the
effective magnetic field first decreases and then increases, 
with a minimum of several
hundred picoseconds around room temperature.

This paper is organized as follows. In Sec.~II, we present the model
and Hamiltonian. In Sec.~III, we study the spin relaxation in rippled graphene based on
the kinetic spin Bloch equations (KSBEs).\cite{wu-review,wu2945,wu373} The effect of
temperature, impurity density and electron density on spin relaxation
is investigated. The anisotropy of spin relaxation, 
without and with a small perpendicular magnetic field, 
is also addressed. We summarize in Sec.~IV.

\section{Model and hamiltonian}

\begin{figure}[th]
  {\includegraphics[width=9cm]{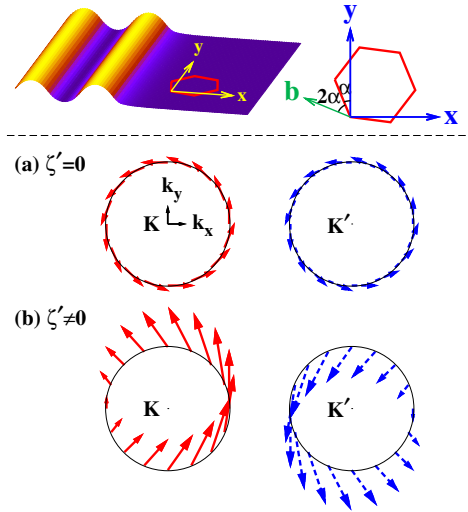}}
  \caption{(Color online) Upper panel: schematic of the  rippled
    graphene curved along the $x$-axis. The angle
    between the $y$-axis and one carbon-carbon bond in counterclockwise
    direction is $\alpha$ ($0\le\alpha<2\pi/3$). Lower panel: schematic of the effective magnetic field
    from the SOC along one circle around ${\bf K}$ and ${\bf
      K}^\prime$ points. (a) The effective magnetic field induced by
    the Rashba-type SOC only ($\zeta^\prime=0$); (b) The effective
    magnetic field from both types of the SOC with $\alpha=0$.}
  \label{figzzw1}
\end{figure} 
The quasi-periodically rippled graphene can be synthesized by chemical vapor
deposition on copper first and then transferred to the SiO$_2$
substrate perpendicular to the $z$-axis.\cite{avsar} The surface
morphology of the rippled graphene is illustrated in the upper panel of Fig.~\ref{figzzw1}. 
The graphene is curved along the $x$-axis and we define the angle
between the $y$-axis and one carbon-carbon bond in counterclockwise
direction as $\alpha$ ($0\le\alpha<2\pi/3$).\cite{jeong} The size and height 
of the ripples are about 50 and 1~nm respectively.\cite{avsar} The radius $R$ of curvature is 
100-200~nm.\cite{avsar} According to Ref.~\onlinecite{jeong}, the local effective SOC
induced by the curvature reads
\begin{equation}
  H_{\rm soc}=\zeta \kappa (\mu\tau_x\otimes
    \sigma_y-\tau_y\otimes \sigma_x)+\mu\zeta^\prime \kappa I\otimes {\bf b}\cdot{\bgreek \sigma},
\label{soc}
\end{equation}
with ${\bf b}=(\cos(\pi/2+3\alpha),\sin(\pi/2+3\alpha),0)$ shown
in the upper panel of Fig.~\ref{figzzw1}. Here ${\bgreek \tau}$ and $I$ are the Pauli matrices and unit
matrix in the pseudospin space. ${\bgreek \sigma}$ are the Pauli
matrices in the spin space. The curvature $\kappa=R^{-1}$. The parameters
$\zeta=0.15$~meV$\cdot$nm and
$\zeta^\prime=0.21$~meV$\cdot$nm.\cite{jeong} On the right-hand side
of Eq.~(\ref{soc}), the first term is the Rashba-type SOC reported previously\cite{Hernando_06} and the
second term is the additional Zeeman-like term diagonal in the pseudospin
space.\cite{jeong}

With the basis laid out in Refs.~\onlinecite{Fabian_SR} and 
\onlinecite{yzhou}, the effective Hamiltonian of the electron band reads
\begin{equation}
    H =\sum_{{\mu}{\bf k}ss^\prime}
  [\varepsilon_{\bf k} \delta_{ss^\prime}
  +(\bm{\Omega}_{\mu\bf k}+\frac{1}{2}g\mu_B{\bf B})\cdot{\bgreek\sigma}_{ss^\prime}] {c_{\mu{\bf k}s}}^{\dagger}
  c_{\mu{\bf k}s^\prime}
  + {H}_{\rm int}.
  \label{H_eff}
\end{equation}
Here $c_{\mu{\bf k}s}$ (${c_{\mu{\bf k}s}}^{\dagger}$) is
the annihilation (creation) operator of electrons in $\mu$ valley
with momentum ${\bf k}$ (relative to the valley center) and spin $s$
($s=\pm \frac{1}{2}$). $\varepsilon_{\bf k}=\hbar v_fk$ with
$v_f=10^6$~m/s. $g\approx 2$ is the effective Land\'e $g$-factor,\cite{zhang136806,lukyanchuk176404} $\mu_B$ is the
Bohr magneton and ${\bf B}$ is an external magnetic field
perpendicular to the graphene plane (when it is applied, its magnitude
is very small and the effect on orbital motion\cite{zhang136806,lukyanchuk176404} is
negligible). The effective magnetic field from the SOC is given by 
\begin{equation}
\bm{\Omega}_{\mu\bf k}=\zeta \kappa (-\sin\theta_{\bf k}, \cos\theta_{\bf
  k},0)+\mu\zeta^\prime \kappa {\bf b},
\label{omega}
\end{equation}
where $\theta_{\bf k}$ is the polar
angle of momentum ${\bf k}$. The second term on the
right-hand side of the equation plays
the role of the Zeeman-like term with the effective static magnetic
field along ${\bf b}$ ($-{\bf b}$) in ${\bf K}$ (${\bf K}^\prime$) 
valley. The effective magnetic fields, without and with
the second term, are schematically plotted in the lower
panel of Fig.~\ref{figzzw1}, where $\alpha$ is set as
zero and hence ${\bf b}$ is along the $y$-aixs. The Hamiltonian $H_{\rm int}$ consists of both the intravalley
and intervalley scatterings. The former include the
electron-impurity,\cite{Adam_08} electron-remote-interfacial
phonon,\cite{fratini} electron-acoustic phonon,\cite{hwang1}
electron--${\bf\Gamma}$-E$_{2g}$ optical phonon\cite{lazzeri} as 
well as electron-electron Coulomb scatterings.\cite{yzhou} The latter include
the electron--${\bf K}$-A$_1^\prime$ optical phonon\cite{lazzeri} and
electron-electron Coulomb scatterings.\cite{yzhou}

\section{Spin relaxation in rippled graphene}
The KSBEs\cite{wu2945,wu373,wu-review} are utilized to study the spin 
relaxation in grahene,\cite{yzhou,zhangdpey}
\begin{equation}
  \partial_t \rho_{\mu\bf k}(t)=\partial_t\rho_{\mu\bf k}(t)|_{\rm coh}+\partial_t\rho_{\mu\bf k}(t)|_{\rm  scat},
\label{ksbee}
\end{equation}
where $\rho_{\mu\bf k}(t)$ represent the density matrices of
electrons with relative momentum ${\bf k}$ in valley ${\mu}$ at
 time $t$. The coherent terms read $\partial_t\rho_{\mu\bf k}(t)|_{\rm
   coh}=-\frac{i}{\hbar}[(\bm{\Omega}_{\mu\bf
   k}+\frac{1}{2}g\mu_B{\bf B})\cdot{\bgreek\sigma},\rho_{\mu {\bf k}}(t)]$, where the Hartree-Fock term from the Coulomb
interaction is neglected due to the small spin
polarization.\cite{wu-review,yzhou,zhangdpey} The concrete expressions of
the scattering terms $\partial_t\rho_{\mu\bf k}(t)|_{\rm  scat}$
are given in Ref.~\onlinecite{yzhou}. By solving the KSBEs, one
obtains the spin relaxation time $\tau_s$ along direction ${\bf n}$ from the time evolution of spin polarization
$P(t)=\frac{1}{n_e}\sum_{\mu {\bf k}}\mbox{Tr}[\rho_{\mu{\bf
    k}}(t){\bgreek \sigma}\cdot{\bf n}]$. In our calculation,
  unless otherwise specified, the initial spin polarization is
  $P(0)=10~\%$, the spin-polarization direction ${\bf n}$ is along the
  $z$-axis, the curvature $\kappa=0.01$~nm$^{-1}$, the electron
  density $n_e=7\times 10^{11}$~cm$^{-2}$, the impurity density
  $n_i=2\times 10^{12}$~cm$^{-2}$ and the external magnetic field
  ${\bf B}=0$. For spin relaxation along the
  $z$-axis, the direction of ${\bf b}$, determined by angle $\alpha$, is irrelevant.

\subsection{Temperature dependence of spin relaxation}
\begin{figure}[ht]
  {\includegraphics[width=7.5cm,height=6.5cm]{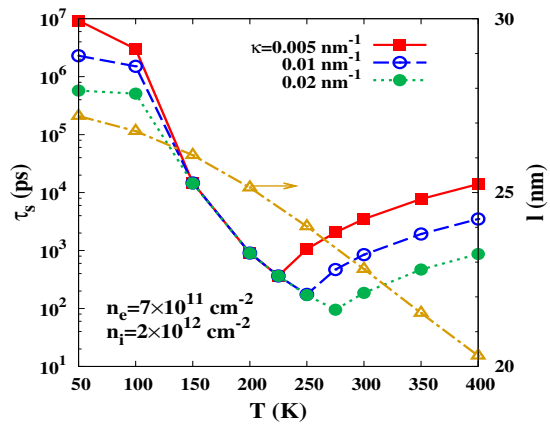}}
  \caption{(Color online) Solid, dashed and dotted curves:
    temperature dependence of spin relaxation
    time with different values of $\kappa$. Chain curve with the scale
    on the right-hand side of the frame: temperature dependence of mean
    free path. $n_e=7\times 10^{11}$~cm$^{-2}$ and $n_i=2\times 10^{12}$~cm$^{-2}$.}
  \label{figzzw2}
\end{figure}

We first study the temperature dependence of the spin relaxation. In
Fig.~\ref{figzzw2}, the spin relaxation time
$\tau_s$ is plotted against temperature $T$ at different curvatures $\kappa$. The electron mean free
path $l$ is around 25~nm (the corresponding mobility is around
$2.6\times 10^3$~cm$^2$/V$\cdot$s) in the whole temperature
regime investigated, as shown in Fig.~\ref{figzzw2} 
with the scale on the right-hand side of the frame. Therefore the
electron mean free path is always smaller than the ripple size. It is indicated by
this figure that when $T\le 100$~K, $\tau_s$ is of the order of
microseconds. However, when $T$ goes beyond 100~K, $\tau_s$ decreases
rapidly to several hundred picoseconds at $T\sim
200$-$300$~K. Nevertheless, when $T$ further 
increases and exceeds the room temperature, $\tau_s$ begins to
increase with $T$. It is also seen from the figure that $\tau_s$ reaches its 
minimum at a higher
temperature with larger curvature $\kappa$. Moreover, when $T=50$~K and $T>250$~K, $\tau_s$ is
proportional to $\kappa^{-2}$. However, in the intermediate
temperature regime $100<T<250$~K, the spin relaxation times are nearly
the same for different values of $\kappa$. This scenario is understood as follows.

We first focus on the case with $\kappa=0.01$~nm$^{-1}$. When $T\le
100$~K, the intervalley scattering is negligible\cite{yzhou} and spins
in two valleys relax independently. The spin relaxation is then determined by the
Rashba-type SOC. The Zeeman-like term only serves as an in-plane
effective magnetic field which mixes the in-plane and out-of-plane spin
relaxations.\cite{dohrmann147405} It is known that with
the Rashba-type SOC only and strong electron-impurity scattering,
$\tau_\perp=\tau_{||}/2=\hbar^2/(4\zeta^2\kappa^2\tau_p)$, where
$\tau_\perp$ ($\tau_{||}$) is the out-of-plane (in-plane) spin relaxation
time.\cite{Fabian_SR,zhanggraphene} Therefore in the presence of the
effective magnetic field, the spin relaxation time along the $z$-axis
is $\tau_s=2/(\tau_\perp^{-1}+\tau_{||}^{-1})=\hbar^2/(3\zeta^2\kappa^2\tau_p)$. With
$\tau_p=l/v_f$, one can estimate $\tau_s\approx 2.4~\mu$s at $T=50$~K, as shown in the
figure. When $T$ increases, the intervalley electron--${\bf
  K}$-A$_1^\prime$ optical phonon scattering becomes important and 
opens another spin relaxation channel together with the opposite effective magnetic
fields in two valleys. According to the model presented in the
  introduction, in the weak intervalley scattering limit
  $\omega\tau_v\ge 1$ ($\omega=2\zeta^\prime\kappa/\hbar$ for the
  concrete situation here), one has the solution 
\begin{equation}
\sum_\mu{\bf  S}_\mu(t)=\frac{2{\bf
    S}_0e^{-t/\tau_v}}{\sqrt{1-(\omega\tau_v)^{-2}}}\sin(\sqrt{\omega^2-\tau_v^{-2}}t+\phi)
\label{s2}
\end{equation}
with $\phi=\arctan\sqrt{\omega^2\tau_v^2-1}$ when ${\bf S}_0$ is
  perpendicular to the effective magnetic field. This indicates that the spin
relaxation time is solely determined by the intervalley scattering time,
  i.e., $\tau_s=\tau_v$. Therefore, in the weak intervalley scattering limit, $\tau_s$
  decreases with the enhancement of the intervalley scattering and hence
  the increase of $T$. However, in the strong intervalley
  scattering limit, $\tau_s=2/(\omega^2\tau_v)$  as
  given by Eq.~(\ref{s1}) in the introduction. In such case, $\tau_s$
  increases with the increase of $T$. The crossover from the weak to strong
  intervalley scattering limit with the increase of $T$ is determined by $\tau_v^{-1}\approx
  \omega=2\zeta^\prime\kappa/\hbar$. At this crossover point,
  $\tau_s\approx\tau_v\approx\hbar/(2\zeta^\prime\kappa)$,
  which is estimated to be 170~ps, just as the value shown in the figure. 

  Based on the above analysis, it is understood that in the zero
  intervalley scattering limit $\tau_s=\hbar^2/(4\zeta^2\kappa^2\tau_p)$ and in the strong
  intervalley scattering limit
  $\tau_s=\hbar^2/(2{\zeta^\prime}^2\kappa^2\tau_v)$. In both limits $\tau_s$ is
  proportional to $\kappa^{-2}$. However, in the weak intervalley
  scattering limit with $\tau_\perp^{-1}\ll\tau_v^{-1}\le\omega$,
  $\tau_s$ is determined by the intervalley scattering time $\tau_v$
  and remains insensitive to $\kappa$. Besides, the crossover
  point in the nonmonotonic temperature dependence of spin relaxation
  time moves to a higher temperature with the increase of curvature $\kappa$ as determined by
  the relation $\tau_v^{-1}\approx 2\zeta^\prime\kappa/\hbar$. These
  properties manifest themselves in the curves with different values
  of $\kappa$ in Fig.~\ref{figzzw2}.

\subsection{Effect of intervalley scattering and SOC on spin relaxation}

The intervalley scatterings include both the electron-electron Coulomb and
electron-phonon scatterings.\cite{yzhou} However, the essential role played in
the spin relaxation channel revealed in this work is the
 intervalley electron-phonon scattering. That is because the
intervalley Coulomb scattering which transfers electrons between 
the two valleys is negligible due to the large
momentum transfer between them and hence 
the small scattering matrix element. Only
the intervalley Coulomb scattering which does not lead to any electron
transfer between the valleys is considered.\cite{yzhou} 

\begin{figure}[ht]
  {\includegraphics[width=7cm]{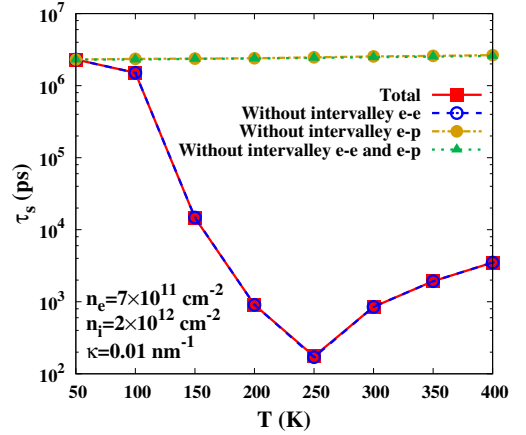}}
  \caption{(Color online) Temperature dependence of spin relaxation
    time with the inclusion of different intervalley scatterings. 
Solid curve: with both the intervalley
    electron-electron (e-e) and electron-phonon (e-p) scatterings;
    Dashed curve: without the intervalley e-e scattering; Chain curve:
    without the intervalley e-p scattering; Dotted curve: without both
    the intervalley e-e and e-p scatterings. $n_e=7\times
    10^{11}$~cm$^{-2}$, $n_i=2\times 10^{12}$~cm$^{-2}$ and $\kappa=0.01$~nm$^{-1}$.}
  \label{figzzw3}
\end{figure}

For comparison, we show the temperature dependence of spin
relaxation time with different intervalley scatterings
included in Fig.~\ref{figzzw3}. It is seen that the intervalley 
Coulomb scattering is unimportant in the whole temperature regime under
study while the intervalley electron-phonon scattering 
affects spin relaxation effectively. Particularly, when the
intervalley electron-phonon scattering is excluded, $\tau_s$ becomes
insensitive to $T$. That is because in the nearly isolated valleys, 
$\tau_s=\hbar^2/(3\zeta^2\kappa^2\tau_p)$ with $\tau_p$ being
dominated by the electron-impurity scattering.\cite{yzhou,zhangdpey}

\begin{figure}[ht]
  {\includegraphics[width=7cm]{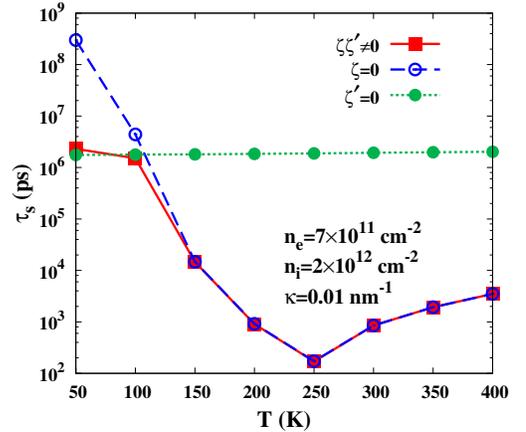}}
  \caption{(Color online) Temperature dependence of spin relaxation
    time for the genuine case ($\zeta\zeta^\prime\neq 0$), the case without the Zeeman-like
    term ($\zeta^\prime=0$) and the case without the Rashba-type SOC
    ($\zeta=0$). $n_e=7\times 10^{11}$~cm$^{-2}$, $n_i=2\times 10^{12}$~cm$^{-2}$ and $\kappa=0.01$~nm$^{-1}$.}
  \label{figzzw4}
\end{figure}

In Fig.~\ref{figzzw4}, we further compare the genuine spin relaxation
($\zeta\zeta^\prime\neq 0$) to the ones without the Zeeman-like
term ($\zeta^\prime=0$) and without the Rashba-type SOC ($\zeta=0$),
in order to reveal the effect of the two types of SOC on spin relaxation. When the
Zeeman-like term is absent ($\zeta^\prime=0$), the two valleys
are degenerate and $\tau_s=\tau_\perp=\hbar^2/(4\zeta^2\kappa^2\tau_p)$.\cite{Fabian_SR,zhanggraphene}
The dotted curve in Fig.~\ref{figzzw4} satisfies this relation very well. 
When the Rashba-type SOC is absent ($\zeta=0$), the spin relaxation is caused by the opposite
effective magnetic fields jointly with the scattering between two
valleys. At low temperature, $\tau_s$ approaches infinity due to the
suppression of intervalley  scattering. By comparing the two
curves with $\zeta=0$ and $\zeta^\prime=0$ to the genuine one with
$\zeta\zeta^\prime\neq 0$, one finds that in the genuine situation the spin relaxation is
dominated by the Rashba-type SOC when $T\le 100$~K and by the Zeeman-like term when $T>100$~K.

\subsection{Spin relaxation with different impurity and electron densities}
\begin{figure}[ht]
  {\includegraphics[width=7cm]{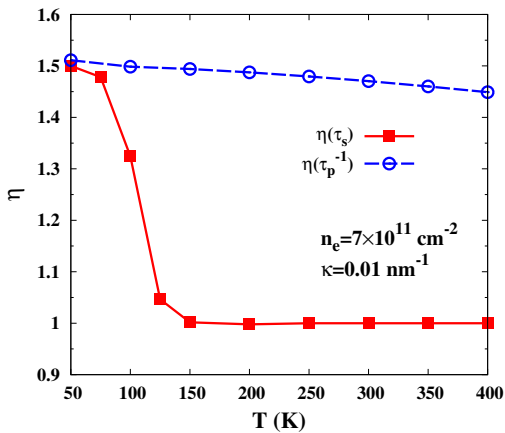}}
  \caption{(Color online) Temperature dependence of the ratios of
    the momentum scattering rate (dashed curve) and the spin relaxation
    time (solid curve) with $n_i=3\times 10^{12}$~cm$^{-2}$ to those with $n_i=2\times
    10^{12}$~cm$^{-2}$, labeled as $\eta(\tau_p^{-1})$ and $\eta(\tau_s)$
    respectively. $n_e=7\times 10^{11}$~cm$^{-2}$ and $\kappa=0.01$~nm$^{-1}$.}
  \label{figzzw5}
\end{figure}

We calculate the spin relaxation with a higher impurity
density $n_i=3\times 10^{12}$~cm$^{-2}$ to explore the effect of
impurity density on spin relaxation. With this impurity density the electron mean
free path $l$ is decreased (compared to the values shown in
Fig.~\ref{figzzw2}) and the condition $l\ll\xi$ is satisfied. In Fig.~\ref{figzzw5}, the
ratios of the momentum scattering rate and spin relaxation time
 with $n_i=3\times 10^{12}$~cm$^{-2}$ to those with $n_i=2\times
 10^{12}$~cm$^{-2}$, labeled as $\eta(\tau_p^{-1})$ and $\eta(\tau_s)$,
are plotted against the temperature. It is
shown that $\eta(\tau_p^{-1})$ remains around 1.5 in the whole temperature
regime under study, as the electron-impurity scattering dominates
 $\tau_p$. $\eta(\tau_s)$ is about 1.5 at $T=50$~K and
rapidly decreases to 1 with the increase of $T$. That is because the spin relaxation is sensitive to
the intravalley scattering only when $\tau_v^{-1}\lesssim\tau_\perp^{-1}$ ($T<150$~K). Particularly, at $T=50$~K,
$\tau_s\propto\tau_p^{-1}$ and
$\eta(\tau_s)=\eta(\tau_p^{-1})=1.5$. When $T\ge 150$~K, the spin relaxation becomes insensitive to the intravalley
scattering and hence the increase of impurity density.

\begin{figure}[ht]
  {\includegraphics[width=7cm]{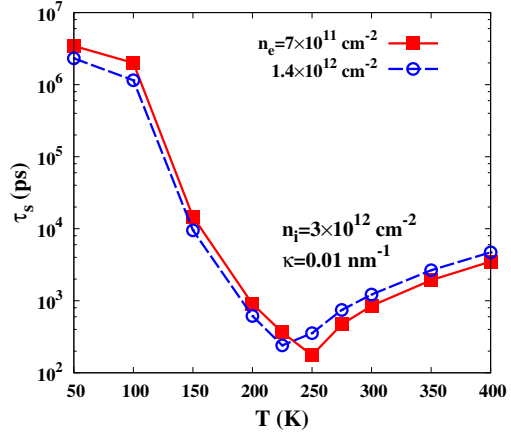}}
  \caption{(Color online) Temperature dependence of spin relaxation
    time with different electron densities. Solid curve: $n_e=7\times
    10^{11}$~cm$^{-2}$; Dashed curve: $n_e=1.4\times
    10^{12}$~cm$^{-2}$. $n_i=3\times 10^{12}$~cm$^{-2}$ and $\kappa=0.01$~nm$^{-1}$.}
  \label{figzzw6}
\end{figure} 

While changing the impurity density influences the spin
relaxation only at low temperature ($T<150$~K), the change of electron
density is expected to affect spin relaxation in a large temperature regime, as
both the intravalley and intervalley scatterings depend on the
electron density. In fact, with the increase of electron density, the intravalley
electron-impurity scattering is weakened while the intervalley
electron-phonon scattering is strengthened. Therefore, with the
increase of electron density, as long as the condition $l\ll \xi$ is
satisfied, the spin relaxation time $\tau_s$ decreases in the weak intervalley scattering
limit ($\tau_s\propto\tau_p^{-1}$ when $T\le 100$~K and
$\tau_s=\tau_v$ when $100<T\le 250$~K) but increases in the strong
intervalley scattering limit ($\tau_s\propto\tau_v^{-1}$ when $T>250$~K). We
calculate the spin relaxation with a higher electron density
$n_e=1.4\times 10^{12}$~cm$^{-2}$ and the impurity density $n_i=3\times
10^{12}$~cm$^{-2}$. For such case the electron mean free path $l$ has the largest value of 30~nm
at $T=50$~K. In Fig.~\ref{figzzw6}, we compare the spin
relaxation for this case to the one with $n_e=7\times 10^{11}$~cm$^{-2}$ and $n_i=3\times
10^{12}$~cm$^{-2}$. It is indeed shown that with the increase of
electron density, $\tau_s$ decreases when $T< 250$~K but increases
when $T\ge$250~K. 

When the electron density is further increased and eventually
$l\ge\xi$, the mechanism discussed in this work is not valid any
more and the spin-flip scattering induced by the fluctuation of the SOC contributes to spin
relaxation.\cite{glazov2157,sherman67,dugaev085306,zhangdpey} When the
latter is dominant, the spin relaxation time $\tau_s$ is of the order of 10~ns and insensitive to
the temperature.\cite{dugaev085306,jeong}

\subsection{Anisotropy of spin relaxation without and with small
  perpendicular magnetic field}

The above studies are limited to the spin relaxation along the
$z$-axis. In fact, due to the in-plane effective magnetic
field along ${\bf b}$, the spin relaxation time along any 
direction perpendicular to
${\bf b}$ is identical. Also, according to the simple model presented in the
introduction, there is no spin relaxation along ${\bf b}$. 
However, due to the Rashba-type
SOC, spins relax along ${\bf b}$ exponentially, with the spin
relaxation time $\tau_s=\tau_{||}$ ($\tau_{||}$ can be obtained from
$\tau_\perp$ given by the dotted curve in Fig.~\ref{figzzw4}, as $\tau_{||}=2\tau_\perp$). This spin relaxation time is of the order of
microseconds and insensitive to temperature. Due to the difference in spin
relaxations along  and  perpendicular to
${\bf b}$, the spin relaxation in the plane parallel to ${\bf b}$,
e.g., the graphene plane, is anisotropic. When the initial spin-polarization
direction ${\bf n}$ deviates from ${\bf b}$ by an angle $\theta_{{\bf
    n},{\bf b}}$, the spin polarization along ${\bf n}$ relaxes in the
form $P(t)=P(0)[\cos^2\theta_{{\bf n},{\bf b}}e^{-t/\tau_{||}}+\sin^2
\theta_{{\bf n},{\bf  b}}f(t)]$. Here 
$f(t)=\frac{e^{-t/\tau_v}}{\sqrt{1-(\omega\tau_v)^{-2}}}
\sin(\sqrt{\omega^2-\tau_v^{-2}}t+\phi)$
in the weak intervalley scattering limit
$\tau_\perp^{-1}\ll\tau_v^{-1}\le\omega$ ($100<T<250$~K) and $f(t)=e^{-\omega^2\tau_vt/2}$
in the strong intervalley scattering limit
$\tau_v^{-1}\gg\omega$ ($T>250$~K), according to Eqs.~(\ref{s2}) and (\ref{s1}) respectively.

\begin{figure}[ht]
  {\includegraphics[width=7cm]{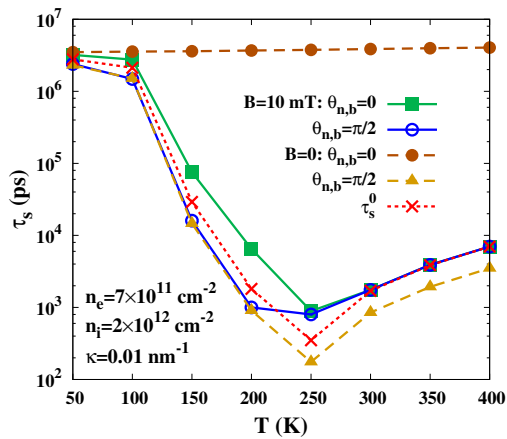}}
  \caption{(Color online) Temperature dependence of spin relaxation
    time along different directions in the graphene plane with (solid curves) and
    without (dashed curves) an external magnetic field along the $z$-axis. Solid
    curve with squares (open circles): spin relaxation along the
    direction parallel (perpendicular) to ${\bf b}$ with 
external magnetic field of
 10~mT along the $z$-axis; Dashed curve with closed circles
    (triangles): spin relaxation along the direction parallel
    (perpendicular) to ${\bf b}$ without  external magnetic field.
 The dotted curve with
    crosses stands for the temperature dependence of the parameter,
    $\tau_s^0$, calculated from the average value of the spin
    relaxation rates along directions parallel and perpendicular to
    ${\bf b}$ without  external magnetic field. 
$n_e=7\times 10^{11}$~cm$^{-2}$, $n_i=2\times
    10^{12}$~cm$^{-2}$ and $\kappa=0.01$~nm$^{-1}$.}
  \label{figzzw7}
\end{figure} 

Usually, the spin relaxation in graphene is experimentally 
studied by the Hanle spin precession 
measurement.\cite{Tombros_08,han222109,Popinciuc,Jozsa_09,han1012.3435,avsar,jo,Pi}
In such measurement, a small magnetic field ${\bf B}$ (at most of the order of 100~mT) perpendicular to the graphene plane
is applied. Under the perpendicular magnetic field, the spin
relaxations along different directions in the graphene plane are
mixed. In the rippled graphene studied here, the spin relaxations along both ${\bf b}$ and the
direction perpendicular to ${\bf b}$ are exponential in the strong intervalley scattering limit
($T>250$~K), with the two spin relaxation rates being $\tau_{||}^{-1}$
and $\omega^2\tau_v/2$, respectively. Therefore, when $B>\hbar|\tau_{||}^{-1}-\omega^2\tau_v/2|/(2g\mu_B)$ (about
10~mT around room temperature), the  spin relaxation along any
direction in the graphene plane has the unique rate
$\tau_s^{-1}=(\tau_{||}^{-1}+\omega^2\tau_v/2)/2\approx\omega^2\tau_v/4$.\cite{dohrmann147405}
However, this is not the case in the weak intervalley scattering limit 
($T<250$~K), as
there is spin oscillation along the direction perpendicular to ${\bf b}$
[refer to Eq.~(\ref{s2})]. We apply a magnetic field of magnitude of 10~mT perpendicular to the
graphene plane and calculate the spin relaxations along ${\bf b}$ ($\theta_{{\bf
    n},{\bf b}}=0$) and the direction perpendicular to ${\bf b}$ in
the graphene plane ($\theta_{{\bf n},{\bf b}}=\pi/2$) respectively at different temperatures. The temperature
dependence of the spin relaxation time along these two
  directions, with and without the perpendicular magnetic field, 
is plotted in Fig.~\ref{figzzw7}. For comparison, we also plot the
 temperature dependence of the parameter, $\tau_s^0$, calculated
from the average value of the perpendicular-magnetic-field--free spin 
relaxation rates along these two directions.
It is shown that due to the perpendicular magnetic field,
the spin relaxation becomes isotropic in the graphene plane when
$T>250$~K, with $\tau_s^{-1}\approx\omega^2\tau_v/4$. However, when
$T<250$~K, the anisotropy is strongly suppressed 
but not completely eliminated. Consequently,
 despite the angle $\alpha$ and the measured
spin-polarization direction, the observed spin relaxation time
 in low-mobility rippled graphene by means of the Hanle spin precession
measurement is expected to have a marked nonmonotonic dependence on
temperature, with a minimum of several hundred picoseconds located
around room temperature.

\section{Conclusion and discussion}
In conclusion, we have studied the electron spin relaxation in rippled
graphene with low mobilities. The electron mean free path is smaller
than the ripple size and the spin relaxation is determined by the
local SOC induced by curvature. The curvature not only leads to the Rashba-type SOC, but also
supplies a Zeeman-like term with opposite effective magnetic fields
along graphene plane in two valleys.\cite{jeong} We show that
this Zeeman-like term, together with the intervalley electron-${\bf K}$-A$_1^\prime$ optical phonon
scattering, gives rise to a spin relaxation channel in rippled
graphene at high temperature (with temperature $T>100$~K). 
This spin relaxation channel can cause a marked nonmonotonic
dependence of spin relaxation on temperature along
 the direction perpendicular to the
effective magnetic field from the Zeeman-like term. A minimal spin
relaxation time of several hundred picoseconds is obtained around room
temperature. However, the Rashba-type SOC dominates the spin relaxation
along the effective magnetic field, leading to a temperature-insensitive
 spin relaxation time of
the order of microseconds.
Therefore, anisotropy exists in the spin relaxation
in the graphene plane. In the presence of a
small perpendicular magnetic field in Hanle spin precession
experiment, the anisotropy of in-plane spin
relaxation is strongly suppressed. Particularly, in the strong
intervalley scattering limit, the spin relaxation in the graphene
plane becomes isotropic, with the spin relaxation time being two times as large as
that along the direction perpendicular to the effective magnetic field
in the absence of the perpendicular magnetic field.

The spin relaxation channel revealed in this work manifests itself only when the electron mean free path is smaller than the
ripple size. However, it is noted
that when this spin relaxation channel is dominant (at temperature
$T>100$~K), the spin relaxation is insensitive to the intravalley
scattering, mainly contributed by the electron-impurity scattering in
the low-mobility sample. It is also noted that in reality the ripples in
graphene may not be quasi-periodic as proposed in this work. The radii of the curvatures may
be spatial dependent.\cite{avsar} Therefore, the observed spin relaxation is coherently averaged
over the curvature morphology in the low-mobility sample.\cite{yzhourandom} In such
case, the spin relaxation is determined by the mean square of curvature in the zero and strong
intervalley scattering limits. The spin relaxation channel revealed in this work should also exist
in carbon nanotude. However, in the work of Semenov {\it et al.} where the 
spin relaxation in carbon nanotube was studied,\cite{semenov} the
intervalley scattering was not considered and this spin relaxation
channel was not incorporated. 

Finally, we discuss the experimentally observed spin relaxation in the
low-mobility rippled graphene by Avsar {\sl et al.}.\cite{avsar} In their sample with an electron density of $7.5\times
10^{11}$~cm$^{-2}$, the in-plane spin relaxation time obtained from the Hanle spin precession
measurement increases mildly from about 130 to 150~ps when $T$
increases from 5 to 300~K.\cite{avsar}  Avsar {\sl et al.} estimated the effect of the
local SOC from curvature and excluded it from the spin
relaxation mechanism (refer to the supplementary information of
Ref.~\onlinecite{avsar}). However, it is noticed that they only took
into account the Rashba-type SOC according to Ref.~\onlinecite{Hernando_06} and the
resulting spin relaxation time is naturally much longer than their
experimental data. The spin relaxation channel revealed
in this work supplies a possible origin of the experimentally observed
short spin relaxation time around room temperature. To account for the
experimental data in the whole temperature regime,
some other factors, e.g., the randomly enhanced SOC by the substrate and/or
adatoms,\cite{Castro_imp,abde,varykhalov,zhanggraphene,zhangdpey,Fabian_SR}
may need to be considered.

\begin{acknowledgments}

This work was supported by  the
National Basic Research Program of China under Grant
No.\,2012CB922002 and the Natural Science Foundation of China
under Grant No.\ 10725417. 
The authors acknowledge valuable discussions with J. S. Jeong.

\end{acknowledgments}

\end{document}